\documentclass[12pt,a4paper,fleqn]{article}
\usepackage{exscale,german,amsthm}
\usepackage{dsfont}
\usepackage[intlimits]{amsmath}
\usepackage[cp850]{inputenc}
\usepackage{epsfig}
\usepackage{stmaryrd}
\usepackage{color}
\usepackage{enumerate}
\usepackage{amssymb}
\usepackage{textcomp}
\usepackage{natbib}
\usepackage{pdflscape}
\newcommand{\bol}[1]{\mbox{\boldmath$#1$}}

\newcommand{\bSigma}{\bol{\Sigma}}

\newcommand{\bmu}{\bol{\mu}}

\newcommand{\bTheta}{\bol{\Theta}}

\newcommand{\btheta}{\bol{\theta}}

\newcommand{\bX}{\mathbf{X}}

\newcommand{\bx}{\mathbf{x}}
\newcommand{\bQ}{\mathbf{Q}}

\newcommand{\bS}{\mathbf{S}}

\newcommand{\bR}{\mathbf{R}}

\newcommand{\bw}{\mathbf{w}}

\newcommand{\bi}{\mathbf{1}}

\newtheorem{theorem}{Theorem}

\newtheorem{corollary}{Corollary}
\newfont{\tabfont}{cmr7 at 7pt}
\begin{document}

\begin{center}
\noindent\textbf{\large Bayesian mean-variance analysis: Optimal portfolio selection under parameter uncertainty}\vspace*{1cm}

\textsc{David Bauder$^a$, Taras Bodnar$^{b,1}$\footnote{$^1$Corresponding Author: Taras Bodnar. E-Mail: taras.bodnar@math.su.se. Tel: +46 8 164562.\\ Fax: +46 8 612 6717.}, Nestor Parolya$^c$, Wolfgang Schmid$^d$}
\vspace*{0.2cm}
{\it \footnotesize \\
$^a$ Department of Mathematics, Humboldt-University of Berlin, D-10099 Berlin, Germany\\
$^b$ Department of Mathematics, Stockholm University, SE-10691 Stockholm, Sweden\\
$^c$ Institute of Statistics, Leibniz University Hannover, D-30167 Hannover, Germany\\
$^d$ Department of Statistics, European University Viadrina, PO Box 1786,  15207 Frankfurt (Oder), Germany
}
\end{center}
\vspace{0.1cm}
 \begin{abstract}
The paper solves the problem of optimal portfolio choice when the parameters of the asset returns distribution, like the mean vector and the covariance matrix are unknown and have to be estimated by using historical data of the asset returns. The new approach employs the Bayesian posterior predictive distribution which is the distribution of the future realization of the asset returns given the observable sample. The parameters of the posterior predictive distributions are functions of the observed data values and, consequently, the solution of the optimization problem is expressed in terms of data only and does not depend on unknown quantities. In contrast, the optimization problem of the traditional approach is based on unknown quantities which are estimated in the second step leading to a suboptimal solution.

We also derive a very useful stochastic representation of the posterior predictive distribution whose application leads not only to the solution of the considered optimization problem, but provides the posterior predictive distribution of the optimal portfolio return used to construct a prediction interval. A Bayesian efficient frontier, a set of optimal portfolios obtained by employing the posterior predictive distribution, is constructed as well. Theoretically and using real data we show that the Bayesian efficient frontier outperforms the sample efficient frontier, a common estimator of the set of optimal portfolios known to be overoptimistic.
\end{abstract}

\vspace{0.2cm}
\noindent Keywords: optimal portfolio, posterior predictive distribution, parameter uncertainty, efficient frontier, stochastic presentation\\[0.2cm]
\noindent JEL Classification: C11, C13, C44, C58, C63 \\


%
\newpage

\section{Introduction}

The fundamental goal of portfolio theory is to allocate optimally the investments between different assets. The mean-variance optimization is a quantitative tool which allows to make this allocation by considering the trade-off between the risk of portfolio and its return. The basic concepts of modern portfolio theory are developed by \cite{Markowitz1952} who introduced a mean-variance portfolio optimization procedure in which investors incorporate their preferences towards the risk and the expected return to seek the best allocation of wealth. This is attained by selecting the portfolios that maximize the expected portfolio return subject to achieving a prespecified level of risk or, equivalently, minimize the variance subject to achieving a prespecified level of expected return. The mean-variance analysis of Markowitz is an important tool for both practitioners and researchers in financial sector today.

The classical problems and pitfalls of the mean-variance analysis are mainly related to extreme weights that often occur when the sample efficient portfolio is constructed. This point was discussed in detail by \cite{merton1980estimating} who presented an estimator of the instantaneous expected return on the market in a log-normal diffusion price model and showed its slow convergence. Moreover, it was proved that the estimates of the variances and of the covariances of the asset returns are more accurate than the estimates of the means. \cite{best1991sensitivity} argued that optimal portfolios are very sensitive to the level of expected returns. Therefore, improving the technique of mean estimation has become a key issue of the portfolio optimization problem recently. The same challenge is also present when the covariance matrix need to be estimated. To this end, \cite{broadie1993computing} showed that the estimated efficient frontier, a set of all mean-variance optimal portfolios overestimates the expected returns of portfolios for different levels of estimation errors. A similar conclusion has also been drawn in more recent studies by \cite{basak2005estimating, siegel2007performance, bodnar2010unbiased}.

An alternative approach to deal with the parameter uncertainty in portfolio analysis is to employ the methods of Bayesian statistics (c.f., \cite{Barry1974}, \cite{Brown1976}, \cite{KleinBawa1976}, \cite{FrostSavarino1986}, \cite{aguilar2000bayesian}, \cite{Rachev2008}, \cite{AvramovZhou2010}, \cite{Sekerke2015}, \cite{BodnarMazurOkhrin2016}). It is remarkable that the Bayesian approach is potentially more attractive since i) it uses prior information about quantities of interest; ii) it facilitates the use of fast, intuitive, and easily implementable numerical algorithms in order to simulate complex economic quantities; iii) it accounts for estimation risk and model uncertainty in the portfolio choice problem. First applications of Bayesian statistics to portfolio analysis during the 1970s were completely based on noninformative or data-based priors. \cite{bawa1979estimation} provided an excellent early survey on such applications. The Bayesian approaches which are based on the diffusion prior are usually comparable with the classical methods for the portfolio selection. However, if some of the risky assets have longer histories than other, then the Bayesian approaches under the diffuse prior lead to different results (see \cite{stambaugh1997analyzing}). \cite{jorion1986bayes} introduced the hyperparameter prior approach in the spirit of the Bayes-Stein shrinkage prior, whereas \cite{black1992global} defended an informal Bayesian analysis with economic arguments and equilibrium relations. They derived the Black-Litterman model which leads to more stable and more diversified portfolios than simple mean-variance optimization. Unfortunately, the application of this model requires a broad variety of data, some of which may be hard to find. Recent studies by \cite{pastor2000portfolio} and \cite{pastor2000comparing} centered prior beliefs around values implied by asset pricing theories. In particular, \cite{pastor2000comparing} investigated the portfolio choices of
 mean-variance-optimizing investors who use sample evidence to update prior beliefs centered on either risk-based or characteristic-based pricing models. \cite{tu2010incorporating} argued that the investment objective provides a useful prior for portfolio selection and proposed an optimal combination of the naive equally weighted portfolio rule with one of the four sophisticated strategies -- the Markowitz rule, the \cite{jorion1986bayes} rule, the \cite{craig2000asset} rule, and the \cite{kan2007optimal} rule -- as a way to improve the performance.

We contribute to the existent literature of optimal portfolio selection by formulating the optimization problem in terms of the posterior predictive distribution and solving it. Using the available information about the development of asset returns which is present in their historical observations, the aim is to construct an optimal portfolio by taking into account investor's preferences. The conventional approach consist of two steps: (i) first, the optimization problem is solved with the solution depending on the unknown parameters of the asset return distribution; (ii) second, the optimal portfolio weights, which are the solutions of optimization problem, are estimated by applying the historical observations of the asset returns. It is important to note that following this approach, the obtained solution is sub-optimal only and it can deviate considerably from the optimal (population) portfolio obtained in the first stage.

In this paper, we propose a new approach, where the solution of the investor's optimization problem is obtained by employing the posterior predictive distribution which takes parameter uncertainty into account before the optimal portfolio choice problem is solved. As a result, its solution is present in terms of historical data and is independent of unknown parameters of the asset return distribution. Consequently, it can be directly applied in practice and, in contrast to the conventional approach, it is optimal.

The rest of the paper is organized as follows. Main theoretical results are given in Section 2. Here, we characterize the posterior predictive distribution of the asset return by developing a very helpful stochastic representation (Theorem 1). This stochastic representation provides not only a way how future realization of portfolio returns could be simulated, but also it is used to derive the first two moments needed in the considered optimization problem. Section 2.2 deals with constructing optimal portfolios by maximizing the posterior mean-variance utility function, while the expression of the Bayesian efficient frontier is derived in Section 2.3. The theoretical results are implement in an empirical study of Section 3, while Section 4 provides a conclusion. The technical derivations are moved to the appendix.

\section{Mean-variance analysis under parameter uncertainty}

\subsection{Posterior predictive distribution}

Let $\bX_t$ denotes the $k$-dimensional vector of returns on asset at time $t$. Assume that a sample of size $n$ of asset returns $\bx_{t-n},...,\bx_{t-1}$, realizations of $\bX_{t-n},...,\bX_{t-1}$, is available which provides the information set $\mathcal{F}_t$ and let $\bx_{(t-1)}=(\bx_{t-n},...,\bx_{t-1})$ be the observation matrix at time $t-1$. Consequently, an investor makes a decision by optimising preferences using information $\mathcal{F}_t$.

Before the decision problem is formulated in Section 2.2, we first derive the predictive posterior distribution $p(\bX_t|\bx_{(t-1)})$ of $\bX_t$ given the previous observation of asset returns summarized in $\bx_{(t-1)}$. The derivation of $p(\bX_t|\bx_{(t-1)})$ is based on the methods of Bayesian statistics which provide well-established techniques for providing inferences of future realizations of asset returns given information $\mathcal{F}_t$.

In the following we assume that the asset returns $\bX_1, \bX_2, ...$ are infinitely exchangeable and multivariate centered spherically symmetric (see, \citet[Section 4.4]{BernardoSmith2000} for the definition and properties). This assumption is very general and it implies that neither the unconditional distribution of the asset returns is normal nor that they are independently distributed. Moreover, the unconditional distribution of the asset returns appears to be heavy-tailed which is usually observed for financial data (see, e.g., \cite{Bradley2003}).

Parameterizing the density function of $\bX_{(t-1)}=(\bX_{t-n},...,\bX_{t-1})$ by the parameter $\btheta$, the posterior distribution of $\btheta$ is obtained by applying the Bayes theorem and it is given by
\begin{eqnarray}\label{post_btheta}
\pi(\btheta|\bx_{(t-1)}) \propto f(\bx_{(t-1)}|\btheta) \pi(\btheta),
\end{eqnarray}
where $\pi(\btheta)$ denotes the prior and $f(\bx_{(t-1)}|\btheta)$ is the likelihood function of $\bX_{(t-1)}$. The posterior distribution $\btheta$ is then used to derive the posterior predictive distribution of the portfolio return at time $t$ expressed as
\begin{equation}\label{exp_ret_port}
X_{p,t}=\bw^\top \bX_t,
\end{equation}
where $\bw=(w_1,...,w_p)^\top$ is the $k$-dimensional vector of portfolio weights.

The posterior distribution \eqref{post_btheta} is employed in the derivation of the posterior predictive distribution as follows:
\begin{eqnarray}\label{post_pred}
f(x_{p,t}|\bx_{(t-1)})=\int_{\btheta \in \bTheta} f(x_{p,t}|\btheta) \pi(\btheta|\bx_{(t-1)}) \mbox{d} \btheta\,.
\end{eqnarray}
Due to the integration present in the definition of the posterior predictive distribution, it is possible to obtaine the analytical expression of $f(x_{p,t}|\bx_{(t-1)})$ only in very rare cases. Moreover, the integration in \eqref{post_pred} could also be high-dimensional, which makes the application of numerical methods very time consuming and also questions the quality of their numerical approximation. In Theorem \ref{th1}, we derive a stochastic representation for the posterior predictive distribution $f(x_{p,t}|\bx_{(t-1)})$ which can be very easily used to draw sample from this distribution as well as to compute its expected value and variance analytically. Finally, it has to be noted that the application of the stochastic representation describing the distribution of random quantities has been used both in the frequentist statistics (see, e.g., \cite{givens2012computational}, \cite{Gupta2013}) and the Bayesian statistics (c.f., \cite{BodnarMazurOkhrin2016}).

\begin{theorem}\label{th1}
Let $\bX_1, \bX_2, ...$ are infinitely exchangeable and multivariate centered spherically symmetric. Let $\pi(\btheta)=|\mathbf{F}|^{1/2}$ be Jeffreys' prior where $|\mathbf{A}|$ denotes the determinant of a square matrix $\mathbf{A}$ and $\mathbf{F}=-\mathbb{E}\left(\frac{\partial^2 \log(f(\bx_{(t-1)}|\btheta))}{\partial \btheta \partial \btheta^\top}\right)$ is the Fisher information matrix. Assume $n>k$. Then the stochastic representation of the random variable $\widehat{X}_{p,t}$ whose density is the posterior predictive distribution \eqref{post_pred} is given by
\begin{eqnarray*}
\widehat{X}_{p,t} &\stackrel{d}{=}& \bw^\top \overline{\bx}_{t-1}
+\sqrt{\bw^\top \bS_{t-1} \bw}\left(\frac{t_1}{\sqrt{n(n-k)}}+ \sqrt{1+\frac{t_1^2}{n-k}}\frac{t_2}{\sqrt{n-k+1}}\right),
\end{eqnarray*}
where
\begin{equation}\label{sample_mean_cov}
\overline{\bx}_{t-1}=\frac{1}{n} \sum_{i=t-n}^{t-1} \bx_i
~~ \text{and} ~~\bS_{t-1}= \sum_{i=t-n}^{t-1}  (\bx_i-\overline{\bx}_{t})(\bx_i-\overline{\bx}_{t})^\top.
\end{equation}
and $t_1$, $t_2$ are independent with $t_1\sim t_{n-k}$ and $t_2\sim t_{n-k+1}$. The symbol ''$\stackrel{d}{=}$'' denotes the equality in distribution.
\end{theorem}

The result of Theorem \ref{th1} provide an easy way how a random sample from the posterior distribution of $f(\bx_t|\bx_{(t-1)})$ can be simulated:
\begin{itemize}
\item[(i)] generate $t_1^{(b)} \sim t_{n-k}$ and $t_2^{(b)}\sim t_{n-k+1}$;
\item[(ii)] compute
\[\widehat{X}_{p,t}^{(b)} = \bw^\top \overline{\bx}_{t}
+\sqrt{\bw^\top \bS_{t} \bw}\left(\frac{t_1^{(b)}}{\sqrt{n(n-k)}}+ \sqrt{1+\frac{(t_1^{(b)})^2}{n-k}}\frac{t_2^{(b)}}{\sqrt{n-k+1}}\right)\]
\item[(iii)] Repeat steps (i) and (ii) for $b=1,...,B$ resulting in independent sample $\widehat{X}_{p,t}^{(1)},...,\widehat{X}_{p,t}^{(B)}$ from the posterior predictive distribution \eqref{post_pred}.
\end{itemize}
The generated sample $\widehat{X}_{p,t}^{(1)},...,\widehat{X}_{p,t}^{(B)}$ is the used to calculate important characteristics of the distribution $f(\bx_t|\bx_{(t-1)})$, like the mean, the variance, the credible interval, etc. To this end, we note that the condition $n>k$ ensures that $\bS_t$ is positive definite and, hence, it is invertible.

Another important application of Theorem \ref{th1} provides us with the analytical expression of the expected value and the variance of the posterior predictive distribution $f(\bx_t|\bx_{(t-1)})$. These findings are formulated in Corollary \ref{cor1}

\begin{corollary}\label{cor1}
Under the conditions of Theorem \ref{th1}, let $n-k>2$. Then:
\begin{equation}\label{post_pred_mean}
\mathbb{E}(\bX_t|\bx_{(t-1)})=\bw^\top \overline{\bx}_{t-1}
\end{equation}
and
\begin{equation}\label{post_pred_mean}
\mathbb{V}ar(\bX_t|\bx_{(t-1)})=c_{k,n}\bw^\top \bS_{t-1} \bw
~~ \text{with} ~~ c_{k,n}=\frac{1}{n-k-1}+\frac{2n-k-1}{n(n-k-1)(n-k-2)}
\end{equation}
\end{corollary}

The proof of Corollary \ref{cor1} is given in the appendix. Its results are used in the next section, where the expressions of optimal portfolio weights are given.

\subsection{Mean-variance optimal portfolios}

The mean-variance investor constructs an optimal portfolio at time $t-1$ for the next period by maximizing the mean-variance utility function given by
\begin{equation}\label{opt}
U(\bw)= \mathbb{E}(\bX_t|\bx_{(t-1)})-\frac{\gamma}{2} \mathbb{V}ar(\bX_t|\bx_{(t-1)})
=\bw^\top \overline{\bx}_{t-1}-\frac{c_{k,n}\gamma}{2}\bw^\top \bS_{t-1} \bw
\end{equation}
under the constraint that the whole wealth is invested into the selected assets, i.e., $\bw^\top \bi=1$ where $\bi$ denotes the $k$-dimensional vector of ones. The quantity $\gamma> 0$ stands for the coefficient of the investor's risk aversion and describes the investor's attitude towards risk.

In contrast to the conventional approach that involves the unknown parameters of the asset return distribution in its formulation, the optimization problem in \eqref{opt} already incorporates the parameter uncertainty by using the available information summarized in the data matrix $\bx_{(t-1)}$. As a result, the output of solving \eqref{opt} is the formula for optimal portfolio weights that could be directly applied in practice, while the estimation of optimal portfolio weights is required in the conventional methods that leads to the suboptimality of the resulting portfolio.

The optimization problem in \eqref{opt} is similar to the optimization problem in the conventional approach (see \cite{ingersoll1987theory,okhrin2006distributional}) with the exception that the risk aversion coefficient is multiplied by the constant $c_{k,n}$. As a results, the solution of \eqref{opt} is given by
\begin{equation}\label{weights_BA}
\bw_{MV,\gamma}=\frac{\bS_{t-1}^{-1}\bi}{\bi^\prime\bS_{t-1}^{-1}\bi}+\gamma^{-1}c_{k,n}^{-1}\bQ_{t-1}\overline{\bx}_{t-1}
~~\text{with}~~
\bQ_{t-1}=\bS_{t-1}^{-1}-\frac{\bS_{t-1}^{-1}\bi\bi^\prime\bS_{t-1}^{-1}}{\bi^\prime\bS_{t-1}^{-1}\bi}
\end{equation}
together with the expected return and the variance expressed as
\begin{equation}\label{mean_BA}
R_{MV,\gamma}=\frac{\bi^\top\bS_{t-1}^{-1}\overline{\bx}_{t-1}}{\bi^\prime\bS_{t-1}^{-1}\bi}+\gamma^{-1}c_{k,n}^{-1}\overline{\bx}_{t-1}^\top\bQ_{t-1}\overline{\bx}_{t-1}
\end{equation}
and
\begin{equation}\label{var_BA}
V_{MV,\gamma}=\frac{c_{k,n}}{\bi^\prime\bS_{t-1}^{-1}\bi}+\gamma^{-2}c_{k,n}^{-1}\overline{\bx}_{t-1}^\top\bQ_{t-1}\overline{\bx}_{t-1},
\end{equation}
respectively, where we use that $\bQ_{t-1} \bi=\mathbf{0}$ and $\bQ_{t-1} \bS_{t-1} \bQ_{t-1} =\bQ_{t-1} $ in \eqref{var_BA}.

Additionally to the formulae of the optimal portfolio weights, the expected return and the variance of the mean-variance optimal portfolios presented in \eqref{weights_BA}-\eqref{var_BA}, the Bayesian approach allows to characterize the posterior predictive distribution of the constructed optimal portfolio. This is achieved by applying the results of Theorem \ref{th1} where the weights of an arbitrary portfolio are replaced by the optimal portfolio weights given in \eqref{weights_BA}. Then, the posterior predictive distribution of the optimal portfolio return is obtained via simulations as described after Theorem \ref{th1} by replacing $\bw$ with $\bw_{MV,\gamma}$ as in \eqref{weights_BA}. This is a very important result which allows the whole characterization of the stochastic behaviour of optimal portfolio return and is a great advantage with respect to the conventional approach where the point estimator is only present.

We conclude this section by noting that the original Markowitz problem (see \cite{Markowitz1952,Markowitz1959}) is solved in the same way. In the mean variance analysis of Markowitz, the optimization problem is given by: (i) minimizing the portfolio variance for a given level of the expected return $R_0$ or (ii) maximizing the expected return for the given level of the variance $V_0$. In the first case the optimal portfolio weights are given by
\begin{equation}\label{weights_BA_Ma1}
\bw_{MV,R_0}=\frac{\bS_{t-1}^{-1}\bi}{\bi^\prime\bS_{t-1}^{-1}\bi}+\left(R_0-\frac{\bi^\top\bS_{t-1}^{-1}\overline{\bx}_{t-1}}{\bi^\prime\bS_{t-1}^{-1}\bi}\right)
\frac{\bQ_{t-1}\overline{\bx}_{t-1}}{\overline{\bx}_{t-1}^\top\bQ_{t-1}\overline{\bx}_{t-1}}
\end{equation}
with
\begin{equation}\label{var_BA_Ma1}
V_{MV,R_0}= \frac{c_{k,n}}{\bi^\prime\bS_{t-1}^{-1}\bi}+c_{k,n}\left(R_0-\frac{\bi^\top\bS_{t-1}^{-1}\overline{\bx}_{t-1}}{\bi^\prime\bS_{t-1}^{-1}\bi}\right)^2
\frac{1}{\overline{\bx}_{t-1}^\top\bQ_{t-1}\overline{\bx}_{t-1}},
\end{equation}
while the solution of the second optimization problem is
\begin{equation}\label{weights_BA_Ma2}
\bw_{MV,V_0}=\frac{\bS_{t-1}^{-1}\bi}{\bi^\prime\bS_{t-1}^{-1}\bi}+\sqrt{c_{k,n}^{-1}V_0-\frac{1}{\bi^\prime\bS_{t-1}^{-1}\bi}}
\frac{\bQ_{t-1}\overline{\bx}_{t-1}}{\sqrt{\overline{\bx}_{t-1}^\top\bQ_{t-1}\overline{\bx}_{t-1}}}
\end{equation}
with
\begin{equation}\label{mean_BA_Ma2}
R_{MV,V_0}=\frac{\bi^\top\bS_{t-1}^{-1}\overline{\bx}_{t-1}}{\bi^\prime\bS_{t-1}^{-1}\bi}+\sqrt{c_{k,n}^{-1}V_0-\frac{1}{\bi^\prime\bS_{t-1}^{-1}\bi}}
\sqrt{\overline{\bx}_{t-1}^\top\bQ_{t-1}\overline{\bx}_{t-1}}.
\end{equation}

\subsection{Bayesian efficient frontier}

Equations \eqref{mean_BA} and \eqref{var_BA} determine the set of all optimal portfolios obtained as solutions of \eqref{opt} for $\gamma >0$. Solving these two equation with respect to $\gamma$ leads to a set in the mean-variance space where all mean-variance optimal portfolios lie. We call this set the Bayesian efficient frontier which is given by
\begin{equation}\label{EF_BA}
\left(R-R_{GMV}\right)^2= \frac{\overline{\bx}_{t-1}^\top\bQ_{t-1}\overline{\bx}_{t-1}}{c_{k,n}}\left(V-V_{GMV}\right),
\end{equation}
where
\begin{equation}\label{GMV}
R_{GMV}=\frac{\bi^\top\bS_{t-1}^{-1}\overline{\bx}_{t-1}}{\bi^\prime\bS_{t-1}^{-1}\bi} ~~\text{and} ~~ V_{GMV}=\frac{c_{k,n}}{\bi^\prime\bS_{t-1}^{-1}\bi}
\end{equation}
are the expected return of the global minimum variance portfolio, i.e., the mean-variance optimal portfolio with the smallest variance, with the weights expressed as
\begin{equation}\label{weights_GMV}
\bw_{GMV}= \frac{\bS_{t-1}^{-1}\bi}{\bi^\prime\bS_{t-1}^{-1}\bi}.
\end{equation}

The quantity $s={\overline{\bx}_{t-1}^\top\bQ_{t-1}\overline{\bx}_{t-1}}/{c_{k,n}}$ is the slope parameter of the efficient frontier which is equal to the amount of the excess squared return with respect to the return of the global minimum variance portfolio when the variance is increased by one. Finally, we note that the Bayesian efficient frontier is a parabola in the mean-variance space which is the same finding as obtained by the conventional approach (see \cite{merton1972analytic}).

\section{Empirical illustration}

\subsection{Data}

For an empirical illustration, we use weekly returns from a collection of assets of the S\&P500, allowing for portfolios ranging from 5 to 40 assets. 
The parameters are estimated with sample sizes of $n\in\{52,78,104,130\}$, corresponding to one year up to two and a half years of weekly data. All the data end on the 8th of October 2017. For $n=52$, this corresponds to almost the whole presidency of Donald Trump, which was, regarding the S\&P500, a period of almost stable growth from 2200 to 2600 points. But besides of two slight drops in August 2015 and the early weeks of 2016, this holds for the other periods - despite of Trump's presidency. The constructed portfolios consist of $k\in\{5,10,25,40\}$ assets. This allows us to analyze the behaviour of the proposed model not only in terms of economic risk but also regarding statistical estimation uncertainty.

\subsection{Conventional approach}
Let $\bmu$ and $\bSigma$ be the mean vector and the covariance matrix of the asset returns. Then the traditional approach to construct an optimal portfolio consists of two steps (see, e.g., \cite{ingersoll1987theory,okhrin2006distributional}):
\begin{itemize}
\item[\textbf{(1)}] The optimization problem
\begin{equation}\label{opt_con}
\bw^\top \bmu-\frac{\gamma}{2}\bw^\top \bSigma \bw \longrightarrow \max ~~\text{subjct to} ~~ \bw^\top \bi=1
\end{equation}
is solved resulting in the expression of optimal portfolio weights presented in terms of the population (unknown) parameters $\bmu$ and $\bSigma$:
\begin{equation}\label{weights_P}
\bw_{P,\gamma}=\frac{\bSigma^{-1}\bi}{\bi^\prime\bSigma^{-1}\bi}+\gamma^{-1}\bR\bmu
~~\text{with}~~
\bR=\bSigma^{-1}-\frac{\bSigma^{-1}\bi\bi^\prime\bSigma^{-1}}{\bi^\prime\bSigma^{-1}\bi}
\end{equation}
with the expected return and the variance expressed as
\begin{equation}\label{mean_var_P}
R_{P,\gamma}=\frac{\bi^\top\bSigma^{-1}\bmu}{\bi^\prime\bSigma^{-1}\bi}+\gamma^{-1}\bmu^\top\bR\bmu
~~ \text{and} ~~
V_{P,\gamma}=\frac{1}{\bi^\prime\bSigma^{-1}\bi}+\gamma^{-2}\bmu^\top\bR\bmu,
\end{equation}

\item[\textbf{(2)}] The unknown population quantities are replaced by their sample counterparts, i.e. by the sample mean vector and the sample covariance matrix given by
\begin{equation*}
\hat{\bmu}=\overline{\bx}_{t-1} ~~ \text{and} ~~ \hat{\bSigma}=d_n\bS_{t-1} ~~ \text{with} ~~ d_n=\frac{1}{n-1}
\end{equation*}

Then the sample optimal portfolio weights are obtained by
\begin{equation}\label{weights_S}
\bw_{S,\gamma}=\frac{\bS_{t-1}^{-1}\bi}{\bi^\prime\bS_{t-1}^{-1}\bi}+\gamma^{-1}d_n^{-1}\bQ_{t-1}\overline{\bx}_{t-1}
\end{equation}
with the sample estimators for the expected return and for the variance given by
\begin{equation}\label{mean_var_S}
R_{S,\gamma}=\frac{\bi^\top\bS_{t-1}^{-1}\overline{\bx}_{t-1}}{\bi^\prime\bS_{t-1}^{-1}\bi}+\gamma^{-1}d_{n}^{-1}\overline{\bx}_{t-1}^\top\bQ_{t-1}\overline{\bx}_{t-1}
~~ \text{and} ~~
V_{S,\gamma}=\frac{d_{n}}{\bi^\prime\bS_{t-1}^{-1}\bi}+\gamma^{-2}d_{n}^{-1}\overline{\bx}_{t-1}^\top\bQ_{t-1}\overline{\bx}_{t-1}\,.
\end{equation}
\end{itemize}

In the similar way, the sample efficient frontier is constructed by (see \cite{bodnar2009estimation,bodnar2009econometrical,kan2008distribution})
\begin{equation}\label{EF_S}
\left(R-R_{GMV,S}\right)^2= \frac{\overline{\bx}_{t-1}^\top\bQ_{t-1}\overline{\bx}_{t-1}}{d_{n}}\left(V-V_{GMV,S}\right),
\end{equation}
where
\begin{equation}\label{GMV_S}
R_{GMV,S}=\frac{\bi^\top\bS_{t-1}^{-1}\overline{\bx}_{t-1}}{\bi^\prime\bS_{t-1}^{-1}\bi} ~~\text{and} ~~ V_{GMV,S}=\frac{d_{n}}{\bi^\prime\bS_{t-1}^{-1}\bi}
\end{equation}
which is an estimator of the population efficient frontier.

\begin{figure}[t]
	\centering
	\includegraphics[width=15cm]{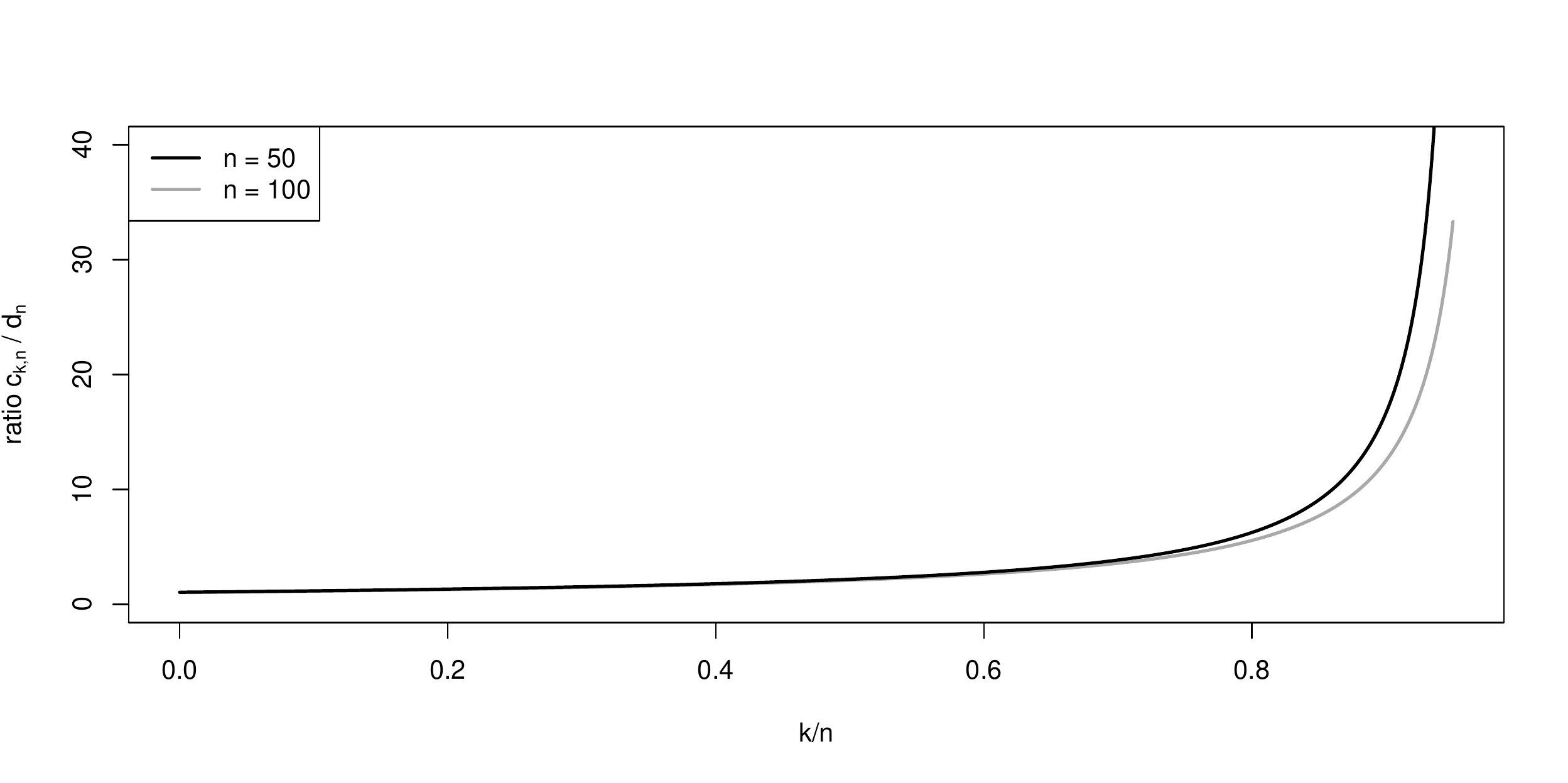}
	\caption{\footnotesize  The ratio $c_{k,n}/d_n$ plotted as a function of $k/n$ for $k/n \in[0,0.95)$ and $n \in \{50, 100\}$. \label{figure1}}
\end{figure}

It is remarkable that the expression of the sample optimal portfolio weights has the same structure as the weights of the optimal portfolios obtained following the Bayesian approach. The only difference is that $c_{k,n}$ in \eqref{weights_BA} is replaced by $d_{n}$ in \eqref{weights_S}. Similar results are also obtained in the case of the efficient frontier which is fully determined by three parameters: the mean and the variance of the global minimum variance portfolio and the slope parameter. While the formulae in the case of the mean of the global minimum variance portfolio coincide, this is not longer true for the variance of the global minimum variance portfolio and the slope coefficient. The Bayesian approach leads to a larger value of the variance and to a smaller value of the slope parameter. The difference between the corresponding expressions obtained by the sample estimation or derived from the Bayeian posterior distribution as in Section 2 can be considerable when the portfolio dimension is comparable to the sample size as shown in Figure \ref{figure1}, where we plot the ratio $c_{k,n}/d_{n}$ as a function of $k/n$ for $n \in\{50,100\}$. We observe that when the number of assets $k$ gets closer to the sample size, even for a moderate ratio of $k/n=0.6$, the Bayesian estimator and the sample estimator deviate. If the number of assets corresponds almost to the sample size, the estimators deviate considerably. Since it is sometimes necessary to restrict an estimation to a smaller sample size, e.g. after a structural break in the data, the difference in the estimators has to be considered.

\begin{figure}[t]
	\centering
	\includegraphics[width=15cm]{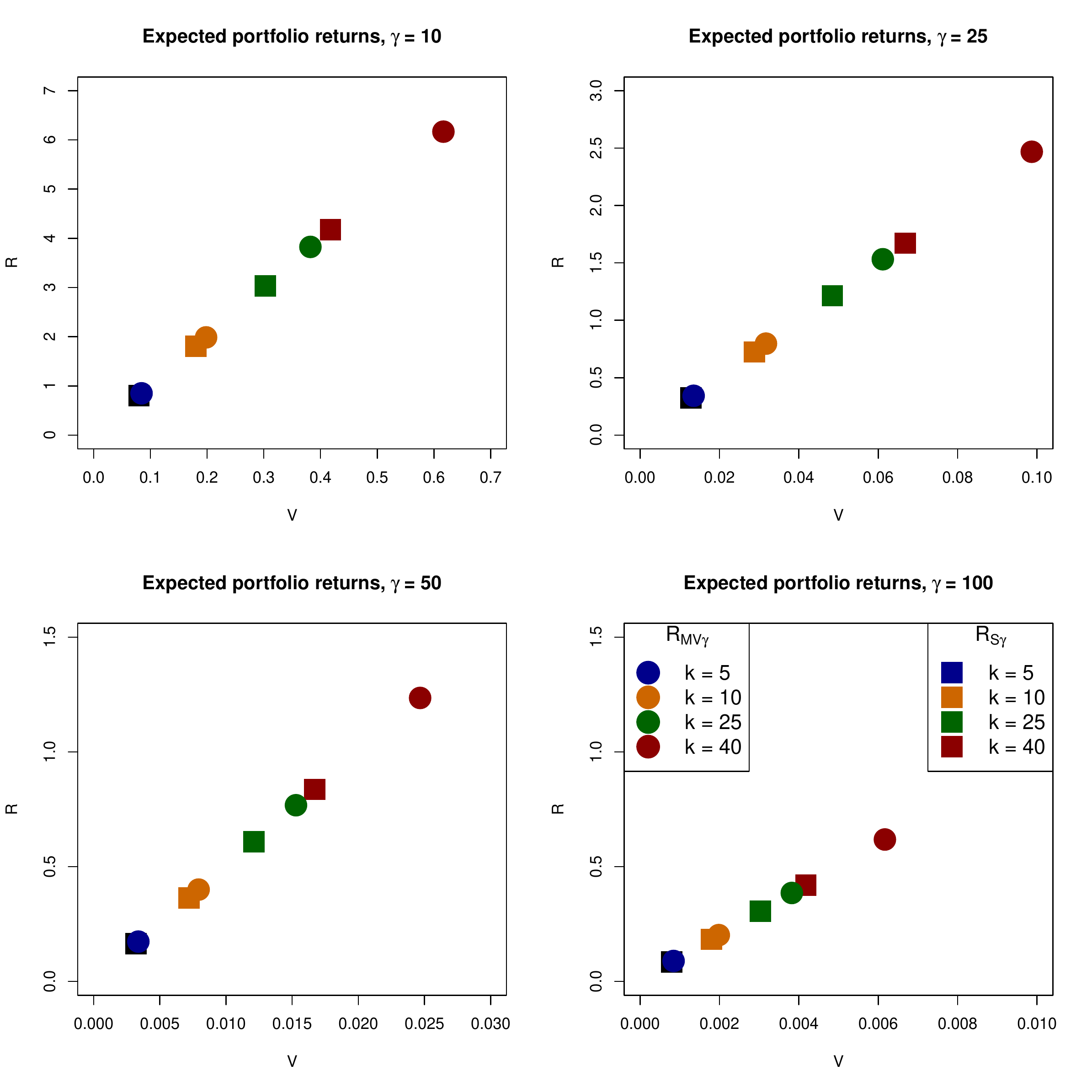}
	\caption{\footnotesize Sample optimal portfolios and Bayesian optimal portfolios for the risk aversion coefficient of $\gamma\in\{10,25,50,100\}$, for the sample case of $n=130$ and for the portfolio dimension of $k\in\{5,10,25,40\}$.  \label{figure2}}
\end{figure}

It is a well-known fact that the sample efficient frontier is overoptimistic and overestimates the location of the population efficient frontier in the mean-variance (c.f., \cite{basak2005estimating, siegel2007performance, bodnar2010unbiased}). In contrast, the Bayesian approach provides an improved procedure which shrinks the sample efficient frontier by increasing the estimated variance of the global minimum portfolio and reducing the slope parameter. We illustrate this point in Section 3.3 on real data described in Section 3.1.

\subsection{Comparison study}

\begin{figure}[t]
	\centering
	\includegraphics[width=15cm]{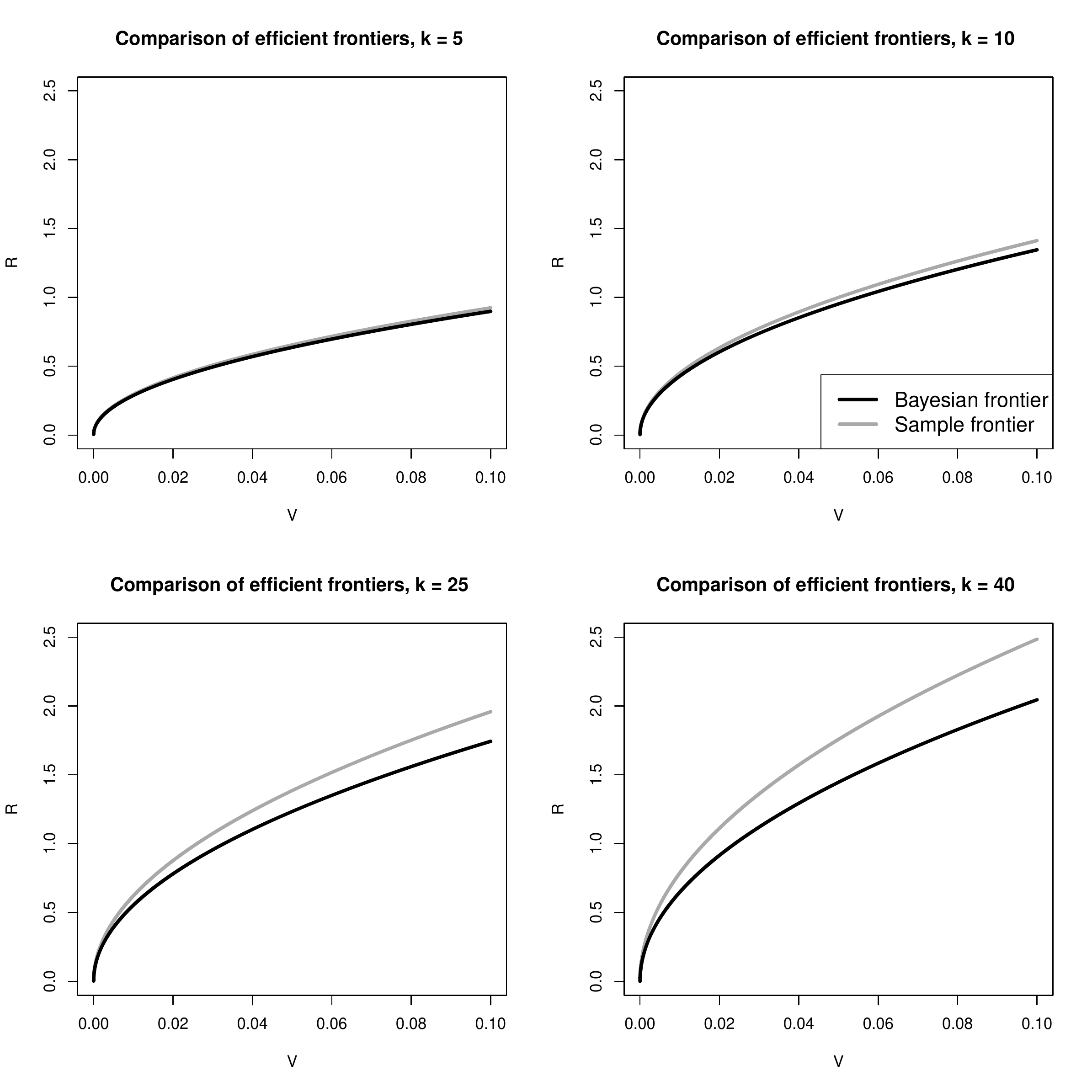}
	\caption{\footnotesize The sample efficient frontiers and the Bayesian efficient frontier for $n=130$ and $k\in\{5,10,25,40\}$. \label{figure3_k}}
\end{figure}

As mentioned in the previous section, there is a distinct difference between the classical sample estimators and the Bayesian estimators proposed in this paper. With this conclusion and the fact that the sample efficient frontier overestimates the population efficient frontier, we expect the estimations for the return and the variance to be larger in the Bayesian case compared to the sample estimations indicating that the Bayesian approach also takes the estimation risk into account in its construction which in practice automatically leads to smaller values of the risk aversion coefficient in comparison to the conventional case. Figure \ref{figure2} illustrates this presumption: fixing $n=130$ and considering different portfolio sizes $k\in\{5,10,25,40\}$ for different risk attitudes $\gamma\in\{10,25,50,100\}$, we find that for the same value of the risk coefficient $\gamma$ and for the same portfolio size, the Bayesian estimator performs as expected compared to the sample estimator. Furthermore, the difference in the estimators increases if the number of assets gets closer to the sample size, as illustrated in Figure \ref{figure1} or when $\gamma$ decreases, i.e. for less risk averse investors the impact of parameter uncertainty becomes larger.

\begin{figure}[t]
	\centering
	\includegraphics[width=15cm]{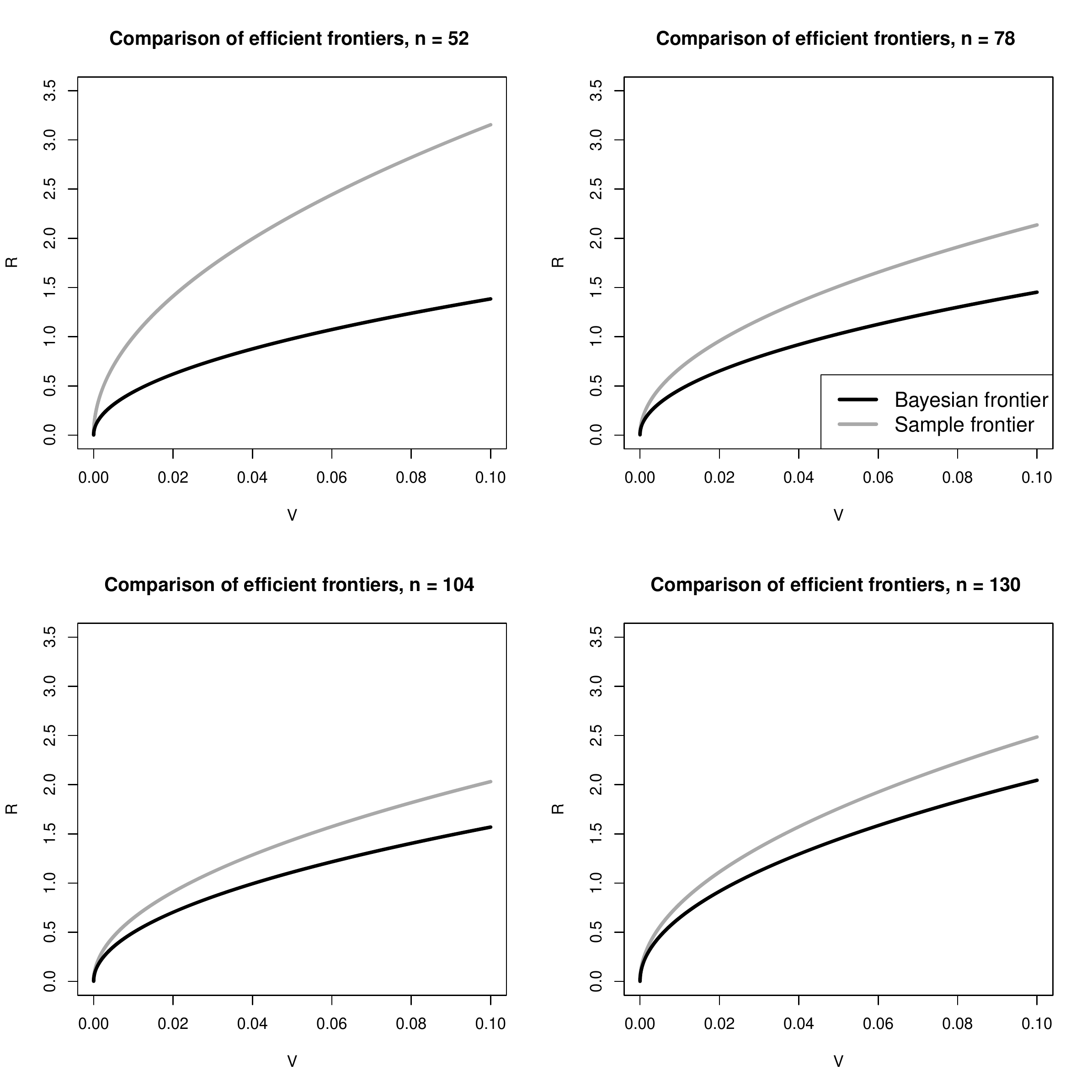}
	\caption{\footnotesize The sample efficient frontiers and the Bayesian efficient frontier for $k=40$ and $n\in\{52,78,104,130\}$. \label{figure3_n}}
\end{figure}

Regarding the efficient frontier, Figure \ref{figure3_k} shows the estimated efficient frontiers for a fixed sample size of $n=130$ and varying portfolio sizes $k\in\{5,10,25,40\}$ in the Bayesian case as well as the conventional case. The Bayesian efficient frontier lies always below the sample efficient frontier and therefore exhibits less overestimation of the population efficient frontier. Furthermore, Figure \ref{figure3_k} also illustrates the finding shown in Figure \ref{figure1}. The estimators of the efficient frontier deviate stronger when the portfolio size gets closer to the sample size. This fact is also illustrated in Figure \ref{figure3_n} for fixed $k=40$ and varying $n\in\{52,78,104,130\}$. The two estimated efficient frontiers coincide more the larger the sample size $n$ is. This is in line with the theoretical implications. Finally, we also observe the increase in the slope parameter of the efficient frontier when the portfolio dimension increases indicating the well-documented positive effect of portfolio diversification.

\subsection{Posterior interval prediction}

\begin{figure}[t]
	\centering
	\includegraphics[width=15cm]{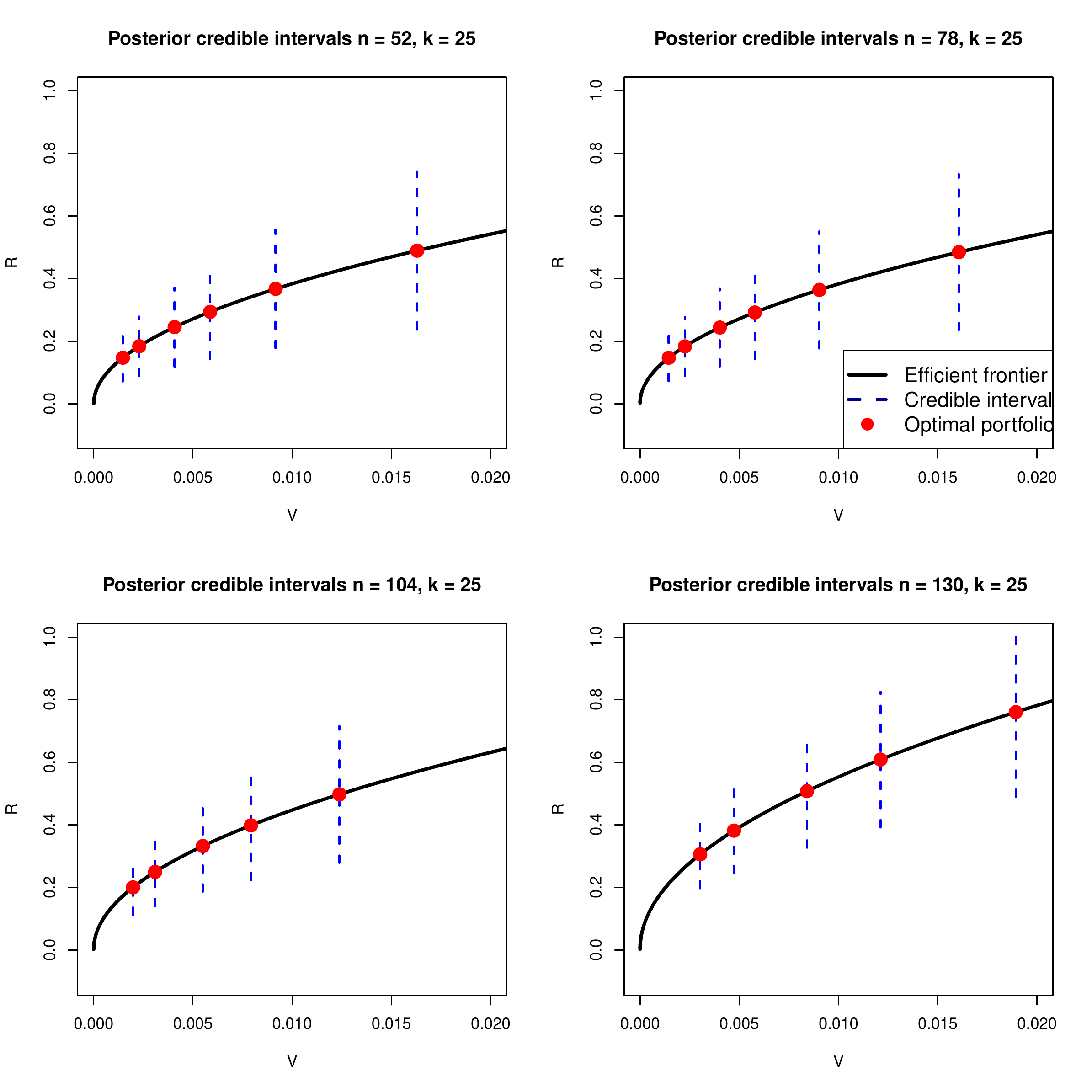}
	\caption{\footnotesize  Credible intervals for the return of optimal portfolios with varying risk attitudes. The sample sizes are chosen to be $n\in\{52,78,104,130\}$ and  the portfolio size is fixed to $k=25$. The confidence level is set to $\alpha=0.05$.  \label{fig4}}
\end{figure}

In contrast to the conventional procedure, the Bayesian approach provides also the whole posterior predictive distribution of the constructed optimal portfolio return and not only the point estimator of its weights. Using data described in Section 3.1, we calculated in this section the prediction intervals for the optimal portfolio returns calculated for several values of the risk-aversion coefficient $\gamma \in \{10,20,...,100\}$, for $k \in \{5,25\}$, and for $n\in\{52,78,104,130\}$ (see Figure \ref{fig4}).

The prediction intervals in Figure \ref{fig4} are obtained as follows:
\begin{itemize}
\item[(a)] Fix $\gamma$ and calculate the expected return and the variance of the corresponding mean-variance optimal portfolio as given \eqref{mean_BA} and \eqref{var_BA};
\item[(b)] For chosen $\gamma$, compute the weights of the optimal mean-variance portfolio $\bw_{MV,\gamma}$ using \eqref{weights_BA}.
\item[(c)] In using $\bw_{MV,\gamma}$ apply the results of Theorem \ref{th1} and the simulation procedure described after the statement of this theorem to get a sample of optimal portfolio returns denoted by $R_{MV,\gamma}^{(b)}$ for $b=1,...,B$.
\item[(d)] Fix the significance level of the prediction interval $\alpha$ and compute the $\alpha/2$- and $(1-\alpha/2)$- quantiles from the empirical distribution of $R_{MV,\gamma}^{(b)}$, $b=1,...,B$
\item[(e)] For the computed value of $V_{GM,\gamma}$ in part (a), plot the point prediction $R_{GM,\gamma}$ from (a) together with the prediction interval from (d).
\end{itemize}

The order of the efficient portfolios given in Figure \ref{fig4} is directly determined by the risk aversion coefficient. The smaller $\gamma$, the riskier is the portfolio and lies therefore more right on the efficient frontier. We observe that the optimal efficient portfolios are shifted to the right for growing sample sizes. But the focus lies here on the credible intervals for a confidence level of $\alpha=0.05$. The first observation is that no credible interval covers negative values, implying positive portfolio returns with probability of $95\%$. The second observation is that the credible intervals become larger the more risky an efficient portfolio becomes -- which is in line with the theory. And the third observation is that these credible intervals for riskier efficient portfolios become larger regardless of the increased sample size. Hence, the decrease in estimation risk resulting from a larger sample is outweighed by the economic risk.

\section{Conclusion}

The mean-variance analysis of Markowitz presents a fundamental way of portfolio construction which is very popular in the financial literature today. It provides an investor the portfolio weights which determine the structure of the optimal portfolio. However, the investor faces with a number of difficulties by implementing this procedure in practice. One of the main pitfalls of the mean-variance analysis is that its solution is presented in terms of unobservable quantities, the parameters of the asset returns distribution. As a results, the optimization problem is performed in two steps. After finding the analytical solution, the optimal portfolio is constructed by replacing the unknown parameters with their estimates. Due to the considerable influence of parameter uncertainty on the investment process, this procedure leads only to sub-optimal portfolios.

We deal with the problem from the viewpoint of Bayesian statistics. The optimization problem is formulated in terms of the posterior predictive distribution which does not involve unknown quantities. Consequently, we deal with parameter uncertainty before solving the optimization problems. This approach allows us to find optimal portfolio weights which now depend only on historical observations of the asset returns. The advantages of the approach are shown both theoretically and empirically. In particular, we show that the constructed Bayesian efficient frontier improves the overoptimism which is present in the sample efficient frontier. Another important advantage of the suggested procedure is that it allows us not only to construct an optimal portfolio based on the posterior predictive distribution, but also an intelligent technique in performing an interval forecast of future realizations of optimal portfolio returns which are obtained by employing the derived stochastic representation of the posterior predictive distribution.

\section*{Appendix}

\begin{proof}[Proof of Theorem 1:]
The assumptions of infinitely exchangeability and multivariate centered spherically symmetry implies (see, e.g., \citet[Proposition 4.6]{BernardoSmith2000}) that the asset returns are independently and identically distributed given the mean vector $\bmu$ and the covariance matrix $\bSigma$ with the conditional distribution given by $\bX_t|\bmu,\bSigma \sim \mathcal{N}_k (\bmu,\bSigma)$ ($k$-dimensional normal distribution with mean vector $\bmu$ and covariance matrix $\bSigma$). Under this model with $\btheta=(\bmu,\bSigma)$, Jeffreys' prior is given by
\begin{eqnarray} \label{Jef_prior}
	\pi(\bmu,\bSigma) &\propto& |\bSigma| ^{-(k+1)/2},
\end{eqnarray}
which leads to the posterior expressed as
\begin{equation}\label{post_mu_sig_d}
\pi(\bmu , \bSigma | \bx_{(t-1)})\propto |\bSigma| ^ {-(n+k+1)/2}\exp \left\{-\frac{n}{2} (\overline{\bx}_{t-1} - \bmu)^\top \bSigma^{-1} (\overline{\bx}_{t-1} - \bmu)
- \frac{1}{2} \mathrm{tr}[\bS_{t-1} \bSigma^{-1}]\right\}\,,
\end{equation}
where $\overline{\bx}_{t-1}$ and $\bS_{t-1}$ are given in the statement of the theorem.

From \eqref{post_mu_sig_d} we obtain that the posterior distribution of $\bSigma$ is the inverse Wishart distribution (see \cite{GuptaNagar2000} for the definition and properties) given by
\begin{equation}\label{cond_post_sig}
\bSigma|\bmu, \bx_{t-1} \sim \mathcal{IW}_k(n+k+1,\widetilde{\bS}_{t-1}(\bmu))~~ \text{with}~~
\widetilde{\bS}_{t-1}(\bmu)=\bS_{t-1}+n(\bmu -\overline{\bx}_{t-1}) (\bmu-\overline{\bx}_{t-1})^\top.
\end{equation}
Furthermore, integrating out $\bSigma$ we get the marginal posterior for $\bmu$ expressed as
\begin{eqnarray*}
\pi (\bmu | \bx_{(t-1)})&\propto& \int_{\bSigma>0} |\bSigma|^{-(n+k+1)/2}\exp \left\{- \frac{1}{2} \mathrm{tr}\left[(n(\overline{\bx}_{t-1} - \bmu) (\overline{\bx}_{t-1} - \bmu)^\top +\bS_{t-1})  \bSigma^{-1}\right]\right\} \mbox{d}\bSigma\\
&\propto& |n(\overline{\bx}_{t-1} - \bmu) (\overline{\bx}_{t-1} - \bmu)^\top +\bS_{t-1}|^{-\frac{n}{2}}\,,
\end{eqnarray*}
where the last equality follows by observing that the function under the integral is the density function of the inverse Wishart distribution with $n+k+1$ degrees of freedom and parameter matrix $n(\overline{\bx}_{t-1}-\bmu) (\overline{\bx}_{t-1} - \bmu)^\top +\bS_{t-1}$. The application of Silvester's determinant theorem leads to
\begin{eqnarray}\label{post_mean}
\pi(\bmu | \bx_{(t-1)})&\propto& \left(1+n(\overline{\bx}_{t-1} - \bmu)^\top \bS_{t-1}^{-1}(\overline{\bx}_{t-1} - \bmu)\right)^{-\frac{n}{2}},
\end{eqnarray}
which proves that $\bmu|\bx_{t-1} \sim t_k\left(n-k, \overline{\bx}_{t-1}, \frac{1}{n(n-k)} \bS_{t-1}\right)$ ($k$-dimensional multivariate $t$-distribution with $n-k$ degrees of freedom, location vector $\overline{\bx}_{t-1}$, and scale matrix $\frac{1}{n(n-k)} \bS_{t-1}$).

Because $\bX_{t-n},...,\bX_t$ are independent given $\bmu$ and $\bSigma$ as well as conditionally normally distributed, we get that the conditional distribution $X_{p,t}|\bmu,\bSigma$ coincides with $X_{p,t}|\bmu,\bSigma,\bx_{(t-1)}$ given by
\begin{eqnarray*}
X_{p,t}|\bmu,\bSigma,\bx_{(t-1)} &\sim& \mathcal{N}(\bw^\top \bmu,\bw^\top \bSigma \bw),
\end{eqnarray*}
where the last equality proves that $X_{p,t}$ depends on $\bmu$, $\bSigma$, and $\bx_{(t-1)}$ only over $\bw^\top \bmu$ and $\bw^\top \bSigma \bw$.

The application of Theorem 3.2.13 in \cite{Muirhead1982} leads to
\begin{equation}\label{th1_eq1}
\frac{\bw^\top \bSigma \bw}{\bw^\top \widetilde{\bS}_{t-1}(\bmu) \bw} \stackrel{d}{=} \frac{1}{\xi},
\end{equation}
where $\xi\sim \chi^2_{n-k+1}$ and is independent of $\bmu$ and $\bX_{(t-1)}$. Then the stochastic representation of $X_{p,t}$ is given by
\begin{eqnarray*}
X_{p,t} &\stackrel{d}{=}& \bw^\top \bmu + \frac{\sqrt{\bw^\top \widetilde{\bS}_{t-1}(\bmu)\bw}}{\sqrt{n-k+1}}t_2\,,
\end{eqnarray*}
where $t_2\sim t_1(n-k+1,0,1)$ is independent of $\bmu$ and $\bX_{(t-1)}$.

Finally, from the properties of the multivariate $t$-distribution, we obtain
\[\bw^\top \bmu- \bw^\top \overline{\bx}_{t-1} \sim t_1\left(n-k,0,\frac{\bw^\top \bS_{t-1} \bw}{n(n-k)} \right)\,,\]
and, consequently,
\begin{equation*}
X_{p,t}\stackrel{d}{=} \bw^\top \overline{\bx}_{t-1}+
\sqrt{\bw^\top \bS_{t-1} \bw}\left(\frac{t_1}{\sqrt{n(n-k)}}+ \sqrt{1+\frac{t_1^2}{n-k}}\frac{t_2}{\sqrt{n-k+1}}\right)\,,
\end{equation*}
where $t_1$ and $t_2$ are independent with $t_1\sim t_{n-k}$ and $t_2\sim t_{n-k+1}$.
\end{proof}

\begin{proof}[Proof of Corollary 1:]
In using the stochastic representation given in Theorem \ref{th1} and the properties of the $t$-distribution, we get
\begin{equation*}
\mathbb{E}(\bX_t|\bx_{(t-1)})=\bw^\top \overline{\bx}_{t-1}+
\sqrt{\bw^\top \bS_{t-1} \bw}\left(\frac{\mathbb{E}(t_1)}{\sqrt{n(n-k)}}+ \mathbb{E}\left(\sqrt{1+\frac{t_1^2}{n-k}}\right)\frac{\mathbb{E}(t_2)}{\sqrt{n-k+1}}\right)=\bw^\top \overline{\bx}_{t-1}
\end{equation*}
and
\begin{eqnarray*}
&&\mathbb{V}ar(\bX_t|\bx_{(t-1)})=\bw^\top \bS_{t-1} \bw \mathbb{V}ar\left(\frac{t_1}{\sqrt{n(n-k)}}+ \sqrt{1+\frac{t_1^2}{n-k}}\frac{t_2}{\sqrt{n-k+1}}\right)\\
&=&\bw^\top \bS_{t-1} \bw \Bigg(\mathbb{E}\left(\frac{t_1^2}{n(n-k)}\right) + \mathbb{E}\left(\left(1+\frac{t_1^2}{n-k}\right)\frac{t_2^2}{n-k+1}\right)\\
&+&2\mathbb{E} \left(\frac{t_1}{\sqrt{n(n-k)}}\sqrt{1+\frac{t_1^2}{n-k}}\frac{t_2}{\sqrt{n-k+1}}\right) \Bigg)\\
&=&\bw^\top \bS_{t-1} \bw \left(\frac{1}{n(n-k)}\mathbb{V}ar(t_1) + \left(1+\frac{1}{n-k}\mathbb{V}ar(t_1)\right)\frac{1}{n-k+1}\mathbb{V}ar(t_2)\right)\\
&=&\bw^\top \bS_{t-1} \bw \Bigg(\frac{1}{n(n-k)}\frac{n-k}{n-k-2} + \left(1+\frac{1}{n-k}\frac{n-k}{n-k-2}\right)\frac{1}{n-k+1}\frac{n-k+1}{n-k-1}\Bigg)\\
&=&\left(\frac{1}{n-k-1}+\frac{2n-k-1}{n(n-k-1)(n-k-2)}\right)\bw^\top \bS_{t-1} \bw \,.
\end{eqnarray*}
\end{proof}

\subsection*{Acknowledgement} This research was partly supported by the German Science Foundation (DFG) via the projects BO 3521/3-1 and SCHM 859/13-1 ''Bayesian Estimation of the Multi-Period Optimal Portfolio Weights and Risk Measures''.

\bibliographystyle{apalike}
\bibliography{MV-BA}

\end{document}